\documentclass[10pt]{iopart}

\eqnobysec
\usepackage{graphicx}
\usepackage{iopams}

\begin{document}

\review[Anomalous transport in low-dimensional]{Anomalous
transport in low-dimensional systems with correlated disorder}

\author{F.~M.~Izrailev\dag\footnote[7]{izrailev@venus.ifuap.buap.mx}\
and N.~M.~Makarov\ddag \footnote[8]{makarov@siu.buap.mx}}

\address{\dag\ Instituto de F\'{\i}sica, Universidad Aut\'{o}noma de Puebla,\\
         Apartado Postal J-48, Puebla, Pue., 72570, M\'{e}xico}

\address{\ddag\ Instituto de Ciencias, Universidad Aut\'{o}noma
         de Puebla, \\ Priv. 17 Norte No. 3417, Col. San Miguel
         Hueyotlipan, Puebla, Pue., 72050, M\'{e}xico}

\begin{abstract}
We review recent results on the anomalous transport in
one-dimensional and quasi-one-dimensional systems with bulk and
surface disorder. Main attention is paid to the role of long-range
correlations in random potentials for the bulk scattering, and in
corrugated profiles for the surface scattering. It is shown that
with a proper choice of correlations one can construct such a
disorder that results in a selective transport with given
properties. A particular interest is in the possibility to arrange
windows of a complete transparency (or reflection) in the
dependence on the wave number of incoming classical waves or
electrons.
\end{abstract}


\submitto{\JPA}

\maketitle

\section{Introduction}

The aim of this paper is two-fold. First, we review recent
developments in the study of low-dimensional models with the
so-called {\it correlated disorder}. By this term we mean specific
long-range correlations embedded in random potentials, that lead
to anomalous transport properties. Second, we observe new results
obtained for one-dimensional (1D) and quasi-1D structures with the
corrugated surfaces resulting in {\it surface scattering}. In the
latter problem, we also consider the case when surface profiles,
although described by random functions, contain the long-range
correlations along the profiles.

Recently, the problem of anomalous transport in 1D systems with
correlated disorder has attracted much attention. For the theory
of disordered systems, the fundamental significance of this
problem is due to exciting results that revise a commonly accepted
belief that any randomness in long 1D structures results in the
Anderson localization. From the experimental viewpoint, many of
the results may have a strong impact for the creation of a new
class of electron nanodevices, optic fibers, acoustic and
electromagnetic waveguides with selective transport properties.

As was shown \cite{IzKr99,IzKrUll01,Lira}, specific long-range
correlations in random potentials can give rise to an appearance
of the {\it mobility edges} for 1D finite structures.
Specifically, an interval of electron energy (or wave frequency)
arises on one side of the mobility edge where the eigenstates turn
out to be extended, in contrast to the other side where the
eigenstates remain to be exponentially localized. The position and
the width of the windows of transparency can be controlled by the
form of the binary correlator of a scattering potential. A quite
simple method was used for constructing random potentials that
result in any predefined energy/frequency window of a perfect
transparency.

The predictions of the theory have been verified experimentally
\cite{KIKS00,KIKS02} by studying transport properties of a
single-mode electromagnetic waveguide with point-like scatters.
The latter were intentionally inserted into the waveguide in order
to provide random potential with a given binary correlator.  In
the experiment, the disorder was created by the array of 500
screws with random heights obtained numerically in accordance with
the analytical expressions. The single-mode regime was provided
within the frequency range $\nu=7.5-15$ GHz. The agreement between
experimental and numerical data was found to be unexpectedly good.
In particular, the predicted windows of a complete reflection
alternated by those of a good transparency, were clearly observed,
in spite of many experimental imperfections.

Main results on the anomalous transport and mobility edges are
based on a new transfer matrix approach
\cite{transmatr,SPIE,Tes02} that was found to be quite effective
in the study of localization properties of eigenstates in 1D
geometry. This approach established a direct relevance of the
Anderson localization to the parametric instability of classical
linear oscillators with white and colored noise \cite{oscill}.
With this approach, the role of long-range correlations has been
substantially studied in 1D discrete tight-binding models of the
Anderson and Kronig-Penney types \cite{IzKr99,IzKrUll01,SPIE}.

A further development of the theory is due to its application to
the surface scattering of electromagnetic waves (or electrons) in
quasi-1D guiding systems. The problem of wave propagation (both
classical and quantum) through such systems with corrugated
surfaces has a quite long history and till now remains a hot topic
in the literature. This problem naturally arises in the analysis
of spectral and transport properties of optics fibers, acoustic
and radio waveguides, remote sensing, shallow water waves,
multilayered systems and photonic lattices, etc.
\cite{BFb79,DeSanBr86}. Also, similar problems arise when
describing the spectral and transport properties of quantum
quasi-particles in thin metal films and semiconductor
nanostructures, such as nanowires and strips, superlattices and
quantum-well-systems
\cite{Chop69,Ber73AR78,LGPb88,BvH90,Land92,FM94,Dat95}.

As is well established, the scattering from corrugated surfaces
results in the classical and quantum diffusive transport
\cite{ChapEnt69}, as well as in the effects of strong
electron/wave localization
\cite{McGM84,BCHMM85,MakYur89FMYu90,TakFer92,MT9801,BulNV96,GTSN98,SFYM9899}.
Correspondingly, the eigenstates of periodic systems with
corrugated surfaces turn out to have a chaotic structure
\cite{LunKyReiKr96}. Recent numerical studies of quasi-1D
surface-disordered systems \cite{GTSN98,SFYM9899} have revealed a
principal difference from those known in the standard models with
{\it bulk} random potentials \cite{FM94}. Specifically, it was
found that transport properties of quasi-1D waveguides with rough
surfaces essentially depend on many characteristic lengths, in
contrast to the bulk scattering where the {\it one-parameter
scaling} occurs. According to this remarkable scaling, all
scattering properties of finite samples with bulk disorder are
determined by the ratio of the localization length found for
infinite systems, to the size of the samples. The situation for
the surface scattering in quasi-1D structures is principally
different due to a non-isotropic character of surface scattering
in the ``channel space''. In particular, the transmission
coefficient for the surface scattering substantially decreases
with an increase of the angle of incoming waves.

One should stress that main results in the theory of surface
scattering were obtained for random surfaces with fast-decaying
correlations along the structures. Therefore, it is of great
importance to explore the role of specific long-range correlations
in surface profiles, having in mind the results found for the 1D
systems with the correlated disorder. Apart from the theoretical
interest, this problem can find many applications since existing
experimental technics allow for the construction of systems with
sophisticated surfaces and bulk-scattering potentials resulting in
anomalous transport properties \cite{WOD95,KS98,B99}.

For single-mode waveguides with the surface scattering the problem
turns out to be equivalent to the 1D bulk scattering. For this
reason, the methods and results obtained for the latter case can
be directly applied for the waveguides
\cite{IzMak01,IzMak02PIERS,IzMak03}. Specifically it was shown,
both analytically and by direct numerical simulations, that
single-mode waveguides with a desired selective transport can be
fabricated by a proper construction of long-range-correlated
random surfaces.

Much more tricky situation arises in the multi-mode waveguides
\cite{IzMak03,IzMak03pss,IzMak04PIERS}. As was shown, in this case
the long-range correlations, on the one hand, give rise to a
suppression of the interaction between different propagating
waveguide modes. On the other hand, the same correlations can
provide a perfect transparency of each independent channel,
similar to what happens in the 1D geometry. The number of
independent transparent modes is governed by the correlation
length and can be as large as the total number of propagating
modes. Therefore, the transmission through waveguide can be
significantly enhanced in comparison with the case of uncorrelated
surface roughness. The important point is that for the multi-mode
surface scattering there are many scattering lengths and the
one-parameter scaling is not valid. As a result, the transport
through different channels can be separated by a proper choice of
long-range correlations. Thus, the intra-mode transitions from
ballistic to diffusive or localized regime are expected to be
sharp enough in order to observe the corresponding mobility edges
experimentally.

In order to understand the impact of long-range correlations in
quasi-1D geometry, we have also discussed the anomalous transport
and mobility edges in waveguides with another kind of disorder,
the so-called, {\it stratified} or {\it layered disorder}
\cite{IzMak04,IzMak04MSMW}. In spite of apparent simplicity of
this model, the results turn out to be quite instructive. This
model describes the scattering of classical waves or electrons
through quasi-1D structures with a disorder that depends on the
longitudinal coordinate only. Therefore, it absorbs both the
properties of 1D correlated structures and those that are specific
for quasi-1D systems with the surface scattering. On the one hand,
the transport through any of open channels is independent from the
others. On the other hand, since the localization length in each
channel strongly depends on its number, the total transmittance is
a complicated combination of the partial 1D transmittances. As a
result, the effect of long-range correlations in such a model
turns out to be highly non-trivial, leading to quite unexpected
phenomena. In particular, it is shown that with a proper choice of
the binary correlator one can arrange the situation when some of
the channels are completely transparent and the others are closed.
This interplay between localized and transparent channels gives
rise to the effect of a {\it non-monotonic} step-wise dependence
of the transmittance on the wave number. The results may find
practical applications for fabrication of electromagnetic/acoustic
waveguides, optic fibers and electron nano-conductors that reveal
non-conventional selective transport.

The structure of this paper is as follows. In next Section 2, the
1D model with a random potential is introduced, for which the main
expressions for the transmission coefficient of finite samples are
given.  With the use of these expressions, we discuss the notion
of the single parameter scaling, as well as the role of long-range
correlations for the anomalous transport. In Section 3 the
quasi-1D model with a stratified disorder is considered in detail.
First, it is shown that in this model an hierarchy of
characteristic scattering lengths arises resulting in a
coexistence of ballistic, diffusive and localized regimes. We
shown how this effect gives rise to unexpected transport
properties for the potentials with specific long-range
correlations. In order to demonstrate these effects, a simple
example is analyzed demonstrating a highly non-trivial dependence
of the transmittance on the energy of incident waves.

Next Section 4 is devoted to the scattering through guiding
systems with corrugated surfaces. In the particular case of a
single-mode waveguides, the analysis of surface scattering is
shown to be analogous to the bulk scattering in 1D structures.
Using this analogy, we show how to construct long-range
correlations along the scattering profiles, in order to observe
the windows of a complete transparency alternated by the regions
of a complete reflection. The analytical results are demonstrated
by a direct numerical computation of the localization length for
the reconstructed potential with a given two-point correlator.
Finally, in the same Section the study of a multi-mode waveguide
with correlated surface profiles is performed, taking as an
example the potential with the step-function for the roughness
power spectrum of a disorder. It is shown that this kind of the
correlations may lead to a fractional dependence of the
transmission coefficient on the normalized wave vector of incoming
waves.

\section{One-dimensional transport in random media}

Let us start with the stationary one-dimensional Schr\"{o}dinger
(or wave) equation
\begin{equation}\label{1DSchr-eq}
\left[\frac{d^{2}}{dx^{2}}+k^{2}-V(x)\right]\Psi(x)=0
\end{equation}
with the wave number $k$ equal to $\omega/c$ for a classical
scalar wave of frequency $\omega$, or to the Fermi wave number for
electrons within the isotropic Fermi-liquid model. The static
potential $V(x)$ for electromagnetic waves is given by
$V(x)=k^{2}\delta\epsilon (x)$ with the dielectric constant
$\epsilon(x)=1-\delta\epsilon(x)$. For electrons, we use the units
in which $2m/\hbar^{2}=1$. Below we assume that the statistically
homogeneous, isotropic random function $V(x)$ is determined by
standard properties,
\begin{eqnarray}\label{V-cor}
\langle V(x)\rangle=0,\qquad\langle V^2(x)\rangle=V_0^2,
\nonumber\\[6pt]
\langle V(x)V(x')\rangle=V_0^2\,{\cal W}(x-x').
\end{eqnarray}
Here the angular brackets $\langle \ldots \rangle $ stand for the
statistical average over different realizations of the random
potential $V(x)$, and the variance is denoted by $V_0^2$. The
binary (two-point) correlator ${\cal W}(x-x')$ is normalized to
${\cal W}(0)=1$ at $x=x'$, and is assumed to decrease with
increasing the distance $|x-x'|$. Obviously, a statistical
treatment is meaningful if the scale of decrease of the correlator
${\cal W}(x)$ is much less than the sample length $L$. Note that
the random potential $V(x)$ is the only source of electron/wave
scattering in the considered model.

In what follows we consider weak over-barrier scattering,
$V_{0}\ll k^{2}$, for which appropriate perturbative approaches
can be used. In this case all transport properties are entirely
determined by the {\it randomness power spectrum} $W(k_x)$,
\numparts
\begin{eqnarray}
{\cal W}(x)&=&\int_{-\infty}^{\infty}\frac{dk_x}{2\pi}
\exp\left(ik_xx\right)\,W(k_x), \label{FR-W} \\[6pt]
W(k_x)&=&\int_{-\infty}^{\infty}dx\exp(-ik_xx)\,{\cal W}(x).
\label{FT-W}
\end{eqnarray}
\endnumparts
Since the correlator ${\cal W}(x)$ is a real and even function of
the coordinate $x$, its Fourier transform (\ref{FT-W}) is even,
real and non-negative function of the wave number $k_x$. The
condition ${\cal W}(0)=1$ results in the following normalization
for $W(k_x)$,
\begin{equation}\label{W-norm}
\int_{-\infty}^{\infty}\frac{dk_x}{2\pi}\,W(k_x)=1.
\end{equation}

We shall characterize the transport properties of finite open
systems by the wave {\it transmittance} or, the same, by the
electron {\it dimensionless conductance} $T$.  In the case of the
electron transport, $T$ is defined as the conductance $G$ in units
of $e^2/\pi\hbar$,
\begin{equation}\label{T-def}
T=\frac{G}{e^2/\pi\hbar}.
\end{equation}
The transmittance $T$ for our model (\ref{1DSchr-eq}) --
(\ref{T-def}) can be found with the use of well developed methods,
such as, e.g., the perturbative diagrammatic technique of
Berezinski \cite{Ber73AR78}, the invariant imbedding method
\cite{BWb75Kb86}, or the two-scale approach \cite{MT9801,MakL99}.
All the methods allow to take adequately into account the effects
of coherent multiple electron/wave scattering giving rise to the
Anderson localization.

\subsection{One-parameter scaling}

Main theoretical result is that the {\it average transmittance}
$\langle T\rangle$ as well as all moments $\langle T^s\rangle$ are
described by the universal function,
\begin{eqnarray}\label{<1D-Ts>}
\langle T^s(L/L_{loc})\rangle&=&\sqrt{\frac{2}{\pi}}
\left(\frac{L_{loc}}{L}\right)^{3/2}
\exp{\left(-\frac{L}{2L_{loc}}\right)}\nonumber\\[6pt]
&\times&\int_0^\infty\frac{zdz}{\cosh^{2s-1}z}
\exp\left(-z^2\frac{L_{loc}}{2L}\right) \int_0^z
dy\cosh^{2(s-1)}y\,,
\nonumber\\[6pt] &&\ s=0,\pm1,\pm2,\ldots\,.
\end{eqnarray}
This function depends solely on the scaling parameter $L/L_{loc}$
which is the ratio between the length $L$ and the so-called {\it
localization length} $L_{loc}$ \cite{LGPb88}. The quantity
$L_{loc}/2$ is, in fact, the {\it backscattering length} in an
infinite sample with the same disorder. The inverse value
$L_{loc}^{-1}$ can be determined via the Lyapunov exponent of the
corresponding averaged wave function $\langle\Psi(x)\rangle$
within the transfer matrix approach. Such a dependence of the
transport properties on the ratio $L/L_{loc}$ manifests a
principal concept of {\it one-parameter scaling} that constitutes
the phenomenon of the 1D Anderson localization.

From equation~(\ref{<1D-Ts>}) one can find relatively easy the low
moments of the distribution of the transmittance $T$.
Specifically, at $s=-1$ one obtains the average dimensionless
resistance $\langle T^{-1}(L/L_{loc})\rangle$. It is given by the
surprisingly simple formula,
\begin{equation} \label{1D-avRes}
\langle T^{-1}(L/L_{loc})\rangle =\frac{1}{2}
\left[1+\exp\left(\frac{4L}{L_{loc}}\right)\right],
\end{equation}
that demonstrates an exponential increase of the resistance with
an increase of the sample length $L$, or with a decrease of the
localization length $L_{loc}$.

The average transmittance (dimensionless conductance) $\langle
T(L/L_{loc})\rangle$ is described by equation~(\ref{<1D-Ts>}) with
$s=1$, leading to the following expression,
\begin{eqnarray}\label{1D-avT}
\langle T(L/L_{loc})\rangle&=&\sqrt{\frac{2}{\pi}}
\left(\frac{L_{loc}}{L}\right)^{3/2}
\exp{\left(-\frac{L}{2L_{loc}}\right)}\nonumber\\[6pt]
&&\times\int_0^\infty\frac{z^2dz}{\cosh z}
\exp\left(-z^2\frac{L_{loc}}{2L}\right).
\end{eqnarray}
As for the second moments, $s=2$ and $s=-2$, they define the
variance of the conductance and resistance, respectively. One can
see that the average resistance does not coincide with the inverse
value of the average conductance, $\langle
T^{-1}\rangle\neq\langle T\rangle^{-1}$. Also, from
equation~(\ref{<1D-Ts>}) one can find that at $L_{loc}\lesssim L$,
the variance of the resistance and conductance is of the order of
their squared average values themselves. This fact means that in
this case both the conductance and resistance are not
self-averaged quantities. Hence, by changing the length $L$ of the
sample, or the disorder itself for the same $L$, one should expect
large fluctuations of the conductance and resistance. This type of
fluctuations are known as the {\it mesoscopic fluctuations} which
are characteristic of strong quantum effects on the macroscopic
scale.

In order to properly characterize the transport properties of
samples with exponentially small conductance, one should refer to
the self-average logarithm of the conductance,
\begin{equation}\label{1D-avLog}
\langle\ln T(L/L_{loc})\rangle=-2L/L_{loc}.
\end{equation}
This result shows an exponential decrease of the transmittance for
the so-called representative (non-resonant) realizations of the
random potential \cite{LGPb88}. Note that this expression can
serve as the definition of the localization length $L_{loc}$
itself, which is complimentary to the definition of $L_{loc}$
through the Lyapunov exponent in the transfer matrix method.

In accordance with the one parameter scaling concept, there are
only {\it two transport regimes} in the 1D disordered structures,
the regimes of the ballistic and localized transport.

{\bf(i)} The {\it ballistic transport} occurs if the localization
length $L_{loc}$ is much larger than the sample length $L$. In
this case the samples are practically fully transparent,
\begin{equation}\label{1D-Tbal}
\langle T(L/L_{loc})\rangle\approx 1-2L/L_{loc}
\qquad\mbox{for}\qquad L_{loc}\gg L.
\end{equation}

{\bf(ii)} Otherwise, the 1D disordered structures exhibit the {\it
localized transport} when the localization length $L_{loc}$ is
small enough in comparison with the sample length $L$. In this
case the average transmittance is exponentially small,
\begin{eqnarray}\label{1D-Tloc}
\langle T(L/L_{loc})\rangle&\approx&\frac{\pi^3}{4\sqrt{2\pi}}
\left(L/L_{loc}\right)^{-3/2}
\exp\left(-L/2L_{loc}\right)\nonumber\\[6pt]
&&\mbox{for}\qquad L_{loc}\ll L.
\end{eqnarray}
As one can see, in the localization regime a 1D disordered system
almost perfectly (with an exponential accuracy) reflects the
electrons (or waves) due to a {\it strong localization} of all
eigenstates.

\subsection{Binary correlator}

The expressions (\ref{<1D-Ts>}) -- (\ref{1D-Tloc}) are universal
and applicable for any one-dimensional system with a {\it weak
static disorder}. As one can see, in order to describe transport
properties of finite sample, one needs to know the localization
length $L_{loc}$ which is a characteristic of the Anderson
localization occurring in infinite samples. According to different
approaches \cite{Ber73AR78,LGPb88,BWb75Kb86,MakL99},  the inverse
localization length for any kind of weak disorder is determined by
the $2k$-harmonic in the randomness power spectrum $W(k_x)$ of the
scattering potential $V(x)$,
\begin{equation}\label{1D-Lloc}
L_{loc}^{-1}(k)=\frac{V_0^2}{8k^2}\,W(2k).
\end{equation}
This canonical expression indicates that all features of the
electron/wave transmission through the 1D disordered media depend
on the two-point correlations in the random scattering potential.
Therefore, if the power spectrum $W(2k)$ is very small or vanishes
within some interval of the wave number $k$, the localization
length $L_{loc}$ turns out to be very large ($L_{loc}\gg L$) or
even diverges. Evidently, the localization effects can be
neglected in this case, and the conductor even of a large length
should be almost fully transparent. This means that, in principle,
by a proper choice of the disorder one can design the disordered
structures with selective ({\it anomalous}) ballistic transport
within a prescribed range of $k$.

Thus, the important practical problem arises of how to construct
such a random potential $V(x)$ from a prescribed power spectrum
$W(k_x)$. This problem can be solved by a direct generalization of
the method developed in the references~\cite{IzKr99,IzKrUll01} for
discrete 1D models with random potentials. In our case of the {\it
continuous} random potential $V(x)$, the solution of the problem
is as follows. First, starting from a given form of the power
spectrum $W(k_x)$, one should obtain the function $\beta(x)$ whose
Fourier transform is $W^{1/2}(k_x)$,
\begin{equation}\label{beta-def}
\beta(x)=\int_{-\infty}^{\infty}\frac{dk_x}{2\pi}
\exp\left(ik_xx\right)\,W^{1/2}(k_x).
\end{equation}
After this, the random profile of the scattering potential $V(x)$
can be generated as a convolution of a delta-correlated random
process $Z(x)$ with the function $\beta(x)$,
\begin{equation}\label{V-beta}
V(x)=V_0\,\int_{-\infty}^\infty\,dx'\,Z(x-x')\,\beta(x').
\end{equation}
Here, the ``white noise'' $Z(x)$ is determined by the standard
properties,
\begin{equation}\label{Zcorr}
\langle Z(x)\rangle=0, \qquad\qquad \langle
Z(x)Z(x')\rangle=\delta(x-x'),
\end{equation}
and can be easily created with the use of random number
generators.

The above expressions allow us to solve the inverse scattering
problem of constructing random potentials from their power
spectrum. Note that this construction is possible in the case of a
weak disorder only. That is why only the binary correlator is
involved in the reconstruction of $V(x)$ while the higher
correlators do not contribute. Note also that the potential
obtained by the proposed method is not unique. Indeed, there is an
infinite set of delta-correlated random processes $Z(x)$.
Therefore, substituting them into the definition (\ref{V-beta}),
we come to an infinite number of random potential profiles with a
given power spectrum $W(k_x)$.

The important point in this approach is that one can arrange a
very {\it sharp transition} between ballistic and localized
regimes of electron/wave transport at any prescribed one or more
points on $k$-axis. It is clear that in this case the power
spectrum $W(k_x)$ must abruptly vanish at the prescribed points.
This means that the binary correlator ${\cal W}(x-x')$ has to be a
slowly decaying function of the distance $x-x'$. In other words,
random scattering potentials $V(x)$ that give rise to the
combination of ballistic and localized transport windows should be
of specific form with long-range correlations along the structure.
Because of an abrupt character, the transition points can be
regarded as {\it mobility edges}.

\subsection{Mobility edges}

Let us now demonstrate the suggested approach of constructing the
potentials giving rise in the mobility edges. For this, we
consider the power spectrum of the following rectangular form
\begin{eqnarray}
W(k_x)&=&\frac{\pi}{2(k_+-k_-)}\,\Theta(2k_+-|k_x|)\,
\Theta(|k_x|-2k_-)\nonumber\\[6pt]
&=&\frac{\pi}{2(k_+-k_-)}\left[\Theta(2k_+-|k_x|)-
\Theta(2k_--|k_x|)\right],\label{WOD-FTW}\\[6pt]
&& k_+>k_->0.\nonumber
\end{eqnarray}
Here $\Theta(x)$ is the Heaviside unit-step function,
$\Theta(x<0)=0$ and $\Theta(x>0)=1$. The characteristic wave
numbers $k_\pm$ are the correlation parameters to be specified.
Note that such a power spectrum was recently employed to create a
special rough surface for the experimental study of an enhanced
backscattering \cite{WOD95}. Moreover, it was used in the
theoretical analysis of light scattering from amplifying media
\cite{SLM01}, as well as of the localization of surface plasmon
polaritons on random surfaces \cite{MSLM01}.

According to equation~(\ref{FR-W}) the binary correlator ${\cal
W}(x)$ corresponding to the expression (\ref{WOD-FTW}), has the
following form,
\begin{equation}\label{WOD-W}
{\cal W}(x)= \frac{\sin(2k_+x)-\sin(2k_-x)}{2(k_+-k_-)x}.
\end{equation}
In equations~(\ref{WOD-FTW}) and (\ref{WOD-W}) the factor
$1/2(k_+-k_-)$ stands to provide the normalization requirement
(\ref{W-norm}), or the same, ${\cal W}(0)=1$.

Following the suggested recipe (\ref{beta-def}) -- (\ref{Zcorr}),
one can construct the scattering potential $V(x)$ with the above
correlation properties,
\begin{equation}\label{V-WOD}
V(x)=\frac{V_0}{\sqrt{2\pi}}\,\int_{-\infty}^\infty\,dx'\,Z(x-x')\,
\frac{\sin(2k_+x')-\sin(2k_-x')}{(k_+-k_-)^{1/2}x'}.
\end{equation}
One should stress that the potential (\ref{V-WOD}) with the binary
correlator (\ref{WOD-W}) and power spectrum (\ref{WOD-FTW}) is
substantially different from the widely used delta-correlated
noise, or random processes with fast-decaying Gaussian
correlations. Here, the potential (\ref{V-WOD}) is specified by
two variation parameters, $(2k_+)^{-1}$ and $(2k_-)^{-1}$, and has
long tales in the expression for the two-point correlator
(\ref{WOD-W}). The existence of such tales manifests the
long-range correlations in the potential, originated from the
step-wise discontinuities at the points $k_x=\pm 2k_\pm$ in its
power spectrum (\ref{WOD-FTW}).

From equations (\ref{1D-Lloc}) and (\ref{WOD-FTW}) one can find
the inverse localization length,
\begin{equation}\label{Lloc-WOD}
\frac{1}{L_{loc}(k)}=\frac{\pi V_0^2}{16(k_+-k_-)}
\frac{\Theta(k_+-k)\Theta(k-k_-)}{k^2}.
\end{equation}
As one can see, there are two mobility edges at the points $k=k_-$
and $k=k_+$. The localization length $L_{loc}(k)$ diverges below
the first point, $k=k_-$, and above the second one, $k=k_+$.
Between these points, for $k_-<k<k_+$, the localization length
(\ref{Lloc-WOD}) has finite value and smoothly increases with an
increase of wave number $k$.

Let us now choose the parameters for which the regime of a strong
localization occurs at the upper transition point $k=k_+$. This
automatically provides the strong localization within the whole
interval $k_-<k<k_+$. The condition of such situation reads as
\begin{equation}\label{LocTr-WOD}
\frac{L}{L_{loc}(k_+)}=\frac{\pi V_0^2L}{16k_+^2(k_+-k_-)}\gg1.
\end{equation}
As a result, there are two regions of transparency for the samples
of finite length $L$ with the chosen random potentials. Between
these regions the average transmittance $\langle T\rangle$ is
exponentially small according to the expression (\ref{1D-Tloc}).
Due to this fact, the system exhibits the localized transport
within the interval $k_-<k<k_+$ and the ballistic regime with
$\langle T\rangle=1$ outside this interval. In the experiment one
can observe that with an increase of the wave number $k$, the
perfect transparency below $k=k_-$ abruptly alternates with a
complete reflection, and recovers at $k=k_+$. From
equations~(\ref{Lloc-WOD}) and (\ref{LocTr-WOD}) one can see that
the smaller value $k_+-k_-$ of the reflecting region the smaller
the length $L_{loc}(k)$ and, consequently, the stronger is the
localization within this region. This remarkable fact may find
important applications in creating a new class of random
narrow-band filters or reflectors.

\section{Quasi-one-dimensional stratified structures}

Now we extend our study of the correlated disorder to the
quasi-one-dimensional (quasi-1D) systems. To this end, we consider
a plane waveguide (or conducting electron wire) of width $d$,
stretched along the $x$-axis. The $z$-axis is directed in
transverse to the waveguide direction so that one (lower) edge of
the waveguide is $z=0$ and the other (upper) edge is $z=d$. The
waveguide is assumed to have a {\it stratified} disorder inside
the finite region $|x|<L/2$ of length $L$, which depends on $x$
only. Thus, the scattering is confined within a region of the
$\{x,z\}$-plane defined by
\begin{equation}\label{domain}
-L/2<x<L/2, \qquad\qquad 0\leq z\leq d.
\end{equation}
As a physically plausible model for stratified disorder, we shall
employ the statistically homogeneous and isotropic random
potential $V(x)$ with the correlation properties (\ref{V-cor}) --
(\ref{W-norm}). One should note that in this case the scale of a
decrease of the binary correlator ${\cal W}(x)$ is of the order of
the characteristic scale of stratification.

The wave function obeys the two-dimensional Schr\"{o}dinger
equation
\begin{equation}\label{2DSchr-eq}
\left[\frac{\partial^2}{\partial x^2}+\frac{\partial^2}{\partial
z^2}+k^{2}-V(x)\right]\Psi(x,z)=0.
\end{equation}
This equation is complemented by zero Dirichlet boundary
conditions on both walls of the waveguide,
$\Psi(x,z=0)=\Psi(x,z=d)=0$.

Since the scattering potential $V(x)$ depends only on the
lengthwise coordinate $x$, the waveguide eigenfunction $\Psi(x,z)$
is naturally presented in the canonical form of the normal
waveguide modes,
\begin{equation}\label{2D-NWM}
\Psi(x,z)=\left(\frac{2}{d}\right)^{1/2}\sin\left(\frac{\pi
nz}{d}\right)\,\psi_n(x).
\end{equation}
Its longitudinal mode-component $\psi_n(x)$ is governed by the
quasi-one-dimensional wave equation that directly follows from
equations~(\ref{2DSchr-eq}) and (\ref{2D-NWM}),
\begin{equation}\label{Q1DSchr-eq}
\left[\frac{d^{2}}{dx^{2}}+k_n^{2}-V(x)\right]\psi_n(x)=0.
\end{equation}
Here the wave number $k_n$ is a quantum value of the lengthwise
wave number $k_x$ for $n$-th waveguide mode,
\begin{equation}\label{kn}
k_n=\sqrt{k^2-(\pi n/d)^2}.
\end{equation}
Evidently, the transport properties are contributed only by those
waveguide modes that can propagate along the structure (along
$x$-axis), i.e. have real value of $k_n$. These propagating modes
occupy the so-called conducting channels. As follows from
equation~(\ref{kn}), the total number $N_d$ of propagating modes
is equal to the integer part $[...]$ of the ratio $kd/\pi$,
\begin{equation}\label{Nd}
N_d=[kd/\pi].
\end{equation}
The waveguide modes with the indices $n>N_d$ corresponding to
imaginary values of $k_n$ and not contributing to the
electron/wave transmission, are called the evanescent modes.

From the wave equation (\ref{Q1DSchr-eq}) one can see that since
the stratified disorder does not depend on $z$, there is no
coupling between the propagating modes. Therefore, the stratified
wire represents a set of 1D non-interacting conducting channels.
That is why, in full accordance with the Landauer's concept
\cite{Land92}, the total transmittance $T$ of the stratified
waveguide can be expressed as a sum of independent partial
transmittances $T_n$ corresponding to every $n$-th propagating
mode,
\begin{equation}\label{Q1D-T-tot}
T(L)=\sum_{n=1}^{N_d}T_n(L).
\end{equation}

In such a way we have reduced the transport problem for the
quasi-1D disordered system to the consideration of the transport
through a number of 1D structures with the random potential
$V(x)$.  The scattering inside every of these structures is
described by equation~(\ref{Q1DSchr-eq}), and is entirely
consistent with the phenomenon of the 1D Anderson localization.
Specifically, the average mode-transmittance $\langle T_n\rangle$
is described by the universal expressions (\ref{<1D-Ts>}) --
(\ref{1D-Tloc}). The {\it mode localization length}
$L_{loc}(k_n)$, associated with a specific $n$-th channel, is
determined by equation~(\ref{1D-Lloc}) in which the total wave
number $k$ should be replaced by the lengthwise wave number $k_n$,
\begin{equation}\label{nmode-Lloc}
L_{loc}^{-1}(k_n) =\frac{V_0^2}{8k_n^2}\,W(2k_n).
\end{equation}
Thus, the partial transport in each $n$-th conducting channel
obeys the concept of the {\it one-parameter scaling}. In
particular, for $L_{loc}(k_n)/L\gg1$, the average transmittance
$\langle T_n\rangle$ exhibits the ballistic behavior and the
corresponding $n$-th normal mode is practically transparent, see
equation~(\ref{1D-Tbal}). On the contrary, the transmittance
$\langle T_n\rangle$ is exponentially small in line with
equation~(\ref{1D-Tloc}), when the mode localization length is
much less than the length of the waveguide, $L_{loc}(k_n)/L\ll1$.
This implies strong electron/wave localization in the $n$-th
channel.

\subsection{Coexistence of ballistic and localized transport}

The main feature of the mode localization length $L_{loc}(k_n)$ is
its rather strong dependence on the channel index $n$. One can see
from equation~(\ref{nmode-Lloc}) that the larger $n$ is, the
smaller is the mode localization length $L_{loc}(k_n)$ and,
consequently, the stronger is the coherent scattering within this
mode. This strong dependence is due to the squared wave number
$k_n$ in the denominator of equation~(\ref{nmode-Lloc}).
Evidently, with an increase of the mode index $n$ the value of
$k_n$ decreases. An additional dependence appears because of the
stratification power spectrum $W(2k_n)$. Since the binary
correlator ${\cal W}(x)$ of the random stratification is a
decreasing function of $|x|$, the numerator $W(2k_n)$ increases
with $n$ (note that it is a constant for the delta-correlated
stratification only). Therefore, both the numerator and
denominator contribute in the same direction for the dependence of
$L_{loc}(k_n)$ on $n$. As a result, we arrive at the hierarchy of
mode localization lengths,
\begin{equation}\label{modeLloc-hierarchy}
L_{loc}(k_{N_d})<L_{loc}(k_{N_d-1})<...<L_{loc}(k_2)<L_{loc}(k_1).
\end{equation}
The smallest mode localization length $L_{loc}(k_{N_d})$ belongs
to the highest (last) channel with the mode index $n=N_d$, while
the largest mode localization length $L_{loc}(k_1)$ corresponds to
the lowest (first) channel with $n=1$. Note that similar hierarchy
was also found in the references~\cite{GTSN98,SFYM9899} for the
attenuations lengths in the model of quasi-1D waveguides with
rough surfaces.

Thus, a remarkable phenomenon arises. On the one hand, the concept
of the one-parameter scaling holds for any of $N_d$ conducting
channels whose partial transport is characterized solely by the
ratio $L/L_{loc}(k_n)$. On the other hand, this concept turns out
to be broken for the total waveguide transport. Indeed, due to the
revealed hierarchy (\ref{modeLloc-hierarchy}) of the mode
localization lengths $L_{loc}(k_n)$, the total average
transmittance (\ref{Q1D-T-tot}) depends on the whole set of
scaling parameters $L/L_{loc}(k_n)$. This fact is in contrast with
quasi-1D bulk-disordered models, for which all transport
properties are shown to be described by one parameter only.

The interplay between the hierarchy of the mode localization
lengths $L_{loc}(k_n)$ on the one side and the one-parameter
scaling for every partial mode-transmittance $\langle T_n\rangle$
on the other side, gives rise to a new phenomenon of the
intermediate coexistence regime, in addition to the regimes of
ballistic or localized transport known in the 1D geometry.

{\bf(i)} {\it Ballistic transport:} If the {\it smallest} mode
localization length $L_{loc}(k_{N_d})$ is much {\it larger} than
the scattering region of size $L$, all conducting channels are
open. They have almost unit partial transmittance, $T_n(L)\approx
1$, hence, the stratified waveguide is fully transparent. In this
case the total average transmittance (\ref{Q1D-T-tot}) is equal to
the total number of the propagating modes,
\begin{equation}\label{Q1D-Tbal}
\langle T(L)\rangle\approx N_d\qquad\mbox{for}\qquad
L_{loc}(k_{N_d})\gg L.
\end{equation}

{\bf(ii)} {\it Localized transport:} On the contrary, if the {\it
largest} of the mode localization lengths $L_{loc}(k_1)$ turns out
to be much {\it smaller} than the waveguide length $L$, all the
propagating modes are strongly localized and the waveguide is
non-transparent. Its total average transmittance, in accordance
with equations~(\ref{Q1D-T-tot}) and (\ref{1D-Tloc}), is
exponentially small,
\begin{eqnarray}\label{Q1D-Tloc}
\langle T(L)\rangle&\approx&\frac{\pi^3}{4\sqrt{2\pi}}
\left[L/L_{loc}(k_1)\right]^{-3/2}
\exp\left[-L/2L_{loc}(k_1)\right]\nonumber\\[6pt]
&&\mbox{for}\qquad L_{loc}(k_1)\ll L.
\end{eqnarray}

{\bf(iii)} {\it Coexistence transport:} The intermediate situation
arises when the {\it smallest} localization length
$L_{loc}(k_{N_d})$ of the last (highest) $N_d$-th mode is {\it
smaller}, while the {\it largest} localization length
$L_{loc}(k_1)$ of the first (lowest) mode is {\it larger} than the
waveguide length $L$,
\begin{equation}\label{Q1D-CoTr}
L_{loc}(k_{N_d})\ll L\ll L_{loc}(k_1).
\end{equation}
In this case an interesting phenomenon of the coexistence of
ballistic and localized transport occurs. Namely, while the {\it
lower} modes are in the ballistic regime, the {\it higher} modes
display the strongly localized behavior.

These transport regimes can be observed experimentally when, e.g.,
the stratification is either the random delta-correlated process
of white-noise type with a constant power spectrum
$V_0^2W(k_x)=W_0=const$, or has Gaussian correlations. As a
demonstration, let us consider the stratified waveguide with a
large number of conducting channels,
\begin{equation}\label{Nd-large}
N_d=[kd/\pi]\approx kd/\pi\gg 1,
\end{equation}
and with a random potential $V(x)$ having the widely used Gaussian
correlator,
\numparts
\begin{eqnarray}
{\cal W}(x)&=&\exp\left(-k_0^2x^2\right),\label{Gaus-cor}\\[6pt]
W(k_x)&=&\sqrt{\pi}\,k_0^{-1}\exp\left(-k_x^2/4k_0^2\right).
\label{Gaus-PS}
\end{eqnarray}
\endnumparts
It is convenient to introduce two parameters
\begin{equation}
\alpha=\frac{L}{L_{loc}^{w}(k_1)}=\frac{W_0L}{8k^2},
\qquad\delta=\frac{L_{loc}^{w}(k_{N_d})}{L_{loc}^{w}(k_1)}=
\frac{2\{kd/\pi\}}{(kd/\pi)}\ll1, \label{aplha-delta}
\end{equation}
where $L_{loc}^{w}(k_1)$ and $L_{loc}^{w}(k_{N_d})$ refer,
respectively, to the largest and smallest mode localization
lengths in the limit case of the white-noise potential
($k_0\to\infty$, $\sqrt{\pi}\,V_0^2k_0^{-1}=W_0=const$). Here
$\{kd/\pi\}$ is the fractional part of the mode parameter
$kd/\pi$.

One can find that for the Gaussian correlations (\ref{Gaus-cor}),
(\ref{Gaus-PS}), all propagating modes are strongly localized when
\begin{equation}\label{Q1d-GC-LocTr}
\exp(k^2/k_0^2)\ll\alpha.
\end{equation}
Note that this inequality is stronger than that valid for the
white-noise case, $\alpha\gg 1$.

The intermediate situation with the coexistence transport occurs
when
\begin{equation}\label{Q1D-GC-CoTr}
\delta\exp(\delta k^2/k_0^2)\ll\alpha\ll\exp(k^2/k_0^2).
\end{equation}
Therefore, the longer range $k_0^{-1}$ of the correlated disorder,
the simpler the conditions (\ref{Q1D-GC-CoTr}) of the {\it
coexistence} of ballistic and localized transport.

Finally, the waveguide is almost perfectly transparent in the case
when
\begin{equation}\label{Q1D-GC-BalTr}
\alpha\ll\delta\exp(\delta k^2/k_0^2).
\end{equation}
From this brief analysis one can conclude that the localization is
easily achieved for the white-noise stratification when the
condition $\alpha\gg 1$, weaker than (\ref{Q1d-GC-LocTr}), should
be met. However, at any given value of $\alpha\gg 1$ (fixed values
of the waveguide length $L$, wave number $k$ and the disorder
strength $W_0$) the intermediate or ballistic transport can be
realized by a proper choice of the stratified correlated disorder,
i.e. with a proper choice of the correlation scale $k_0^{-1}$.
From equation~(\ref{Q1D-GC-CoTr}) one can also understand that for
the standard white-noise case, the coexistence of ballistic and
localized modes can be observed under the conditions
$\delta\ll\alpha\ll1$. If $\alpha\ll\delta$, the quasi-1D
structure with the white-noise stratification displays the
ballistic transport.

\subsection{Correlated stratification and step-wise
non-monotonic transmittance}

From the above analysis it becomes clear that in quasi-1D
stratified guiding structures with delta-correlated or Gaussian
correlations, the crossover from the ballistic to localized
transport is realized through the successive localization of
highest propagating modes. Otherwise, if we start from the
localized transport regime, the crossover to the ballistic
transport is realized via the successive opening (delocalization)
of the lowest conducting channels.

The fundamentally different situation arises when random
stratified media have specific long-range correlations. To show
this, we would like to note that the mode localization length
$L_{loc}(k_n)$ of any $n$-th conducting channel is entirely
determined by the stratification power spectrum $W(k_x)$, see
equation~(\ref{nmode-Lloc}). Therefore, if $W(2k_n)$ abruptly
vanishes for some wave numbers $k_n$, then $L_{loc}(k_n)$ diverges
and the corresponding propagating mode appears to be fully
transparent even for a large length of the waveguide.

Let us take the following binary correlator ${\cal W}(x)$ with the
corresponding step-wise power spectrum $W(k_x)$,
\numparts
\begin{eqnarray}
{\cal W}(x)&=&\pi\delta(2k_cx)-\frac{\sin(2k_cx)}{2k_cx},
\label{Q1D-Wcor}\\[6pt]
W(k_x)&=&\frac{\pi}{2k_c}\,\Theta(|k_x|-2k_c).\label{Q1D-PScor}
\end{eqnarray}
\endnumparts
Here $\delta(x)$ is the Dirac delta-function and the
characteristic wave number $k_c>0$ is a correlation parameter to
be specified. Applying the method defined by the equations
(\ref{beta-def}) -- (\ref{Zcorr}), one can find that the profile
of the random stratification having such correlations can be
fabricated with the use of the potential
\begin{equation}\label{Q1D-Vcor}
V(x)=\frac{V_0}{\sqrt{2\pi k_c}}\left[\pi Z(x)-
\int_{-\infty}^{\infty}dx'Z(x-x')\frac{\sin(2k_cx')}{x'}\right].
\end{equation}
This potential has quite sophisticated form. It is a superposition
of the white noise and the long-range correlated potential.

For the case under consideration the inverse value of the mode
localization length (\ref{nmode-Lloc}) takes the following
explicit form
\begin{equation}\label{Q1D-nLloc-cor}
L_{loc}^{-1}(k_n) =\frac{\pi V_0^2}{16k_ck_n^2}\,\Theta(k_n-k_c).
\end{equation}
This expression leads to very interesting conclusions.

{\bf(i)} All {\it low} propagating modes with the lengthwise wave
numbers $k_n$ that exceed the correlation parameter $k_c$
($k_n>k_c$), have finite mode localization length,
\begin{equation}\label{Q1D-Lnlow}
L_{loc}^{-1}(k_n>k_c)=\pi V_0^2/16k_ck_n^2.
\end{equation}
For large enough waveguide length $L\gg16k_ck_1^2/\pi V_0^2$, all
of them are strongly localized. The requirement $k_n>k_c$ implies
that the mode indices $n$ of localized channels are restricted
from above by the condition
\begin{equation}\label{Q1D-Nloc}
n\leq N_{loc}=[(kd/\pi)(1-k_c^2/k^2)^{1/2}]\Theta(k-k_c).
\end{equation}
Therefore, the integer $N_{loc}$ should be regarded as the total
number of the localized and non-transparent modes.

{\bf(ii)} For {\it high} conducting channels with $k_n<k_c$, the
mode localization length diverges,
\begin{equation}\label{Q1D-Lnhigh}
L_{loc}^{-1}(k_n<k_c)=0.
\end{equation}
Therefore, each of the modes with the index $n>N_{loc}$ has unit
partial transmittance $\langle T_n\rangle=1$ and displays the
ballistic transport. Such modes occupy a subset of ballistic and
completely transparent channels.

{\bf(iii)} The value of $N_{loc}$ determines the total number of
localized modes. The total number $N_{bal}$ of ballistic modes is
evidently equal to $N_{bal}=N_d-N_{loc}$. Since the localized
modes do not contribute to the total average waveguide
transmittance (\ref{Q1D-T-tot}), the latter is equal to the number
of completely transparent modes $N_{bal}$ and do not depend on the
waveguide length $L$,
\begin{equation}\label{Q1D-Ttot-cor}
\langle T\rangle=[kd/\pi]-
[\sqrt{(kd/\pi)^2-(k_cd/\pi)^2}\,]\Theta(k-k_c).
\end{equation}
We remind that square brackets stand for the integer part of the
inner expression. It should be emphasized once more that in
contrast to the usual situation related to the hierarchy of mode
localization lengths (\ref{modeLloc-hierarchy}), now the {\it low}
propagating modes are {\it localized} and non-transparent, while
the {\it high} modes are {\it ballistic} with a perfect
transparency.

For the correlation parameter $k_c\ll k$, the number of localized
modes is of the order of $N_d$,
\begin{equation}\label{Q1D-Nloc-kc>k}
N_{loc}\approx[(kd/\pi)(1-k_c^2/2k^2)]\quad\mbox{for}\quad
k_c/k\ll1.
\end{equation}
Consequently, the number of ballistic modes $N_{bal}$ is small, or
there are no such modes at all. Otherwise, if $k_c\to k$, the
integer $N_{loc}$ turns out to be much less than the total number
of propagating modes $N_d$,
\begin{equation}\label{Q1D-Nloc-kctok}
N_{loc}\approx[\sqrt{2}(kd/\pi)(1-k_c/k)^{1/2}]\ll
N_d\quad\mbox{for}\quad 1-k_c/k\ll1.
\end{equation}
Here the number of transparent modes $N_{loc}$ is large. When
$k_c>k_1$, the number $N_{loc}$ vanishes and all modes become
fully transparent, in spite of their scattering by random
stratification. In this case the correlated disorder results in a
perfect transmission of quantum/classical waves. One can see that
the point $k_1=k_c$ at which the wave number of the first mode
$k_1$ is equal to the correlation wave number $k_c$, is, in
essence, the total mobility edge that separates the region of
complete transparency from that where lower modes are localized.

From the above analysis one can conclude that the transmittance
(\ref{Q1D-Ttot-cor}) of quasi-1D multimode structure with a
long-range correlated stratification reveals a quite unexpected
{\it non-monotonic} step-wise dependence on the total wave number
$k$ that is controlled by the correlation parameter $k_c$. An
example of this dependence is shown in figure~1.

\begin{figure}[htb]
\vspace{0.cm}
\hspace{4.0cm}
\begin{center}
\includegraphics[width=4.2in,height=2.4in,angle=0]{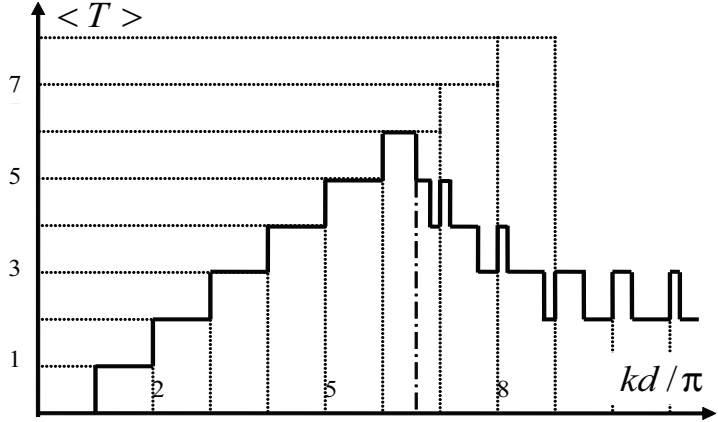}
\vspace{0.0cm} \caption{Non-monotonic step-wise dependence of the
transmittance (\ref{Q1D-Ttot-cor}) of stratified structure versus
the normalized wave number $kd/\pi$ for the dimensionless
correlation parameter $k_cd/\pi=6.5$}
\end{center}
\end{figure}

Within the region where the wave number of the lowest (first) mode
$k_1$ is less than the correlation wave number $k_c$ ($k_1<k_c$
or, the same, $kd/\pi<\sqrt{(k_cd/\pi)^2+1}$), the second term in
the expression (\ref{Q1D-Ttot-cor}) is absent, and all propagating
modes are completely transparent. Here the transmittance exhibits
a ballistic {\it step-wise increase} with an increase of the total
wave number $k$. Each step up arises for an integer value of the
mode parameter $kd/\pi$ when a new conducting channel emerges in
the waveguide. Such step-wise increasing behavior of the total
transmittance is similar to that known to occur in quasi-1D
ballistic {\it non-disordered} structures (see, e.g.,
\cite{vWvHB88}).

With a subsequent increase of $k$, the first-mode wave number
$k_1$ becomes larger than $k_c$ ($k_1\geq k_c$ or
$kd/\pi\geq\sqrt{(k_cd/\pi)^2+1}$). Here the transmittance shows
not only standard steps up associated with the first term in the
equation (\ref{Q1D-Ttot-cor}) but also {\it steps down} that are
originated from the second term. In contrast to the steps up, the
steps down are formed by the correlated scattering and occur due
to successive abrupt localization of low modes at the
corresponding mobility edges. The first step down occurs at the
total mobility edge $k_1=k_c$, where the first mode is localized.
The second step down is due to the mobility edge $k_2=k_c$ of the
second mode, etc. The positions of the steps down are at the
integer values of the square root
$\sqrt{(kd/\pi)^2-(k_cd/\pi)^2}$, see, equation
(\ref{Q1D-Ttot-cor}), and, in general, do not coincide with the
integer values of the mode parameter $kd/\pi$. The interplay
between steps up and down results in a new kind of step-wise {\it
non-monotonic} dependence of the total quasi-1D transmittance.

These results may find practical applications for the fabrication
of electromagnetic/acoustic waveguides, optic fibers and electron
nanodevices with a selective transport. For an abrupt dependence
of the power spectrum at some energy, the transition from the
ballistic to localized transport is expected to be sharp enough in
order to observe them experimentally. For example, for the GaAs
quasi-1D quantum-well structures with the effective electron mass
$m_e=6.7\cdot 10^{-2}m_0$, the Fermi-energy $E_F=7$ meV and
$d=500$ nm, the number of channels is about $17$. Therefore, for
the potential with $\pi/k_c\approx 80$ nm, with an increase of $k$
one can observe about $16$ local mobility edges characterized by
the steps down, after initial $6$ steps up in the conductance
dependence on the wave number $k$.

\section{Surface-corrugated guiding systems}

In this section we consider the so-called surface scattering. This
type of scattering is due to a disorder that arises inherently in
natural atmospheric and oceanic waveguides. Also surface disorder
arises either inherently (due to growth defects, fracture, etc.)
or artificially (e.g., by lithographic preparation ) in guiding
electron/wave nanodevices.

The commonly used model of the surface scattering is a plane
waveguide (or quasi-1D electron wire) of length $L$ and average
width $d$ stretched along the $x$-axis. One (lower) surface of the
waveguide is supposed to be of a rough (corrugated) profile
$z=\xi(x)$, slightly deviated from its flat average $z=0$. The
other (upper) surface is taken, for simplicity, flat, $z=d$. In
other words, the surface-corrugated guiding system occupies the
area defined by the relations,
\begin{equation}\label{SD-region}
-L/2<x<L/2, \qquad\qquad \xi(x)\leq z\leq d.
\end{equation}
The function $\xi(x)$ describing the surface roughness is assumed
to be a statistically homogeneous and isotropic random function
with the following characteristics,
\begin{eqnarray}\label{SD-Ksi}
\langle\xi(x)\rangle=0, \qquad \langle\xi^2(x)\rangle=\sigma^2,
\nonumber\\[6pt]
\langle\xi(x)\xi(x')\rangle=\sigma^2{\cal W}(x-x').
\end{eqnarray}
As in the previous sections, the angular brackets stand for a
statistical averaging over the disorder, i.e. in the case of
surface disorder over different realizations of the random surface
profile $\xi(x)$ with the variance $\sigma^2$. The binary
correlator ${\cal W}(x)$ is normalized to its maximal value,
${\cal W}(0)=1$, and assumed to decrease with increasing $|x|$ on
the scale characterizing the mean length of surface corrugation.
The roughness power spectrum $W(k_x)$ is defined by equations
(\ref{FR-W}) -- (\ref{FT-W}) with the normalization
(\ref{W-norm}). We also note that the surface corrugations are
assumed to be small in height, $\sigma\ll d$, and smooth enough.
These limitations are common in the surface scattering theories
that are based on appropriate perturbative approaches
\cite{BFb79}.

In the $x$-direction the system is open while in the transverse
$z$-direction the zero Dirichlet boundary conditions are applied
at both surfaces, $z=\xi(x)$ and $z=d$. Thus, the analysis of the
surface-scattering in a quasi-1D guiding system is reduced to the
study of the following two-dimensional boundary-value problem,
\begin{eqnarray}\label{2D-BVP}
\left(\frac{\partial^2}{\partial x^2}+\frac{\partial^2}{\partial
z^2}+k^{2}\right)\Psi(x,z)=0,
\nonumber\\[6pt]
\Psi(x,z=\xi(x))=\Psi(x,z=d)=0.
\end{eqnarray}
Comparing with the models studied in the previous sections, here
the Schr\"{o}dinger (or wave) equation does not contain any
scattering potential. In contrast with the bulk scattering, an
electron/wave experiences a perturbation due to roughness of the
lower surface, i.e. due to the nonuniform boundary condition.

\subsection{Single-mode structure}

Keeping in mind the relevance of surface scattering of quantum and
classical waves to the Anderson localization, we, first, consider
a single-mode waveguide. In this case the mode parameter $kd/\pi$
is restricted by the relation $1<kd/\pi<2$ and the number of
conducting channels equals to one, $N_d=1$. The transport through
such systems is provided solely by the lowest normal mode with
$n=1$ that propagates with the lengthwise wave number $k_1$,
\begin{equation}\label{k1}
k_1=\sqrt{k^2-(\pi /d)^2}.
\end{equation}
All other waveguide modes with $n\geq2$ are evanescent and do not
contribute to the transport properties. From the single-mode
condition $1<kd/\pi<2$, it follows that the wave number $k_1$ is
confined within the interval

\begin{equation}\label{SM-k1}
0<k_1d/\pi<\sqrt{3}.
\end{equation}
Note that the assumed weak surface-scattering condition $\sigma\ll
d$ leads to the inequality $k_1\sigma\ll 1$.

As was shown in the papers~\cite{MakYur89FMYu90}, the transport
problem (\ref{2D-BVP}) for the surface-disordered single-mode
waveguide corresponds to that for the 1D disordered model
(\ref{1DSchr-eq}) with the lengthwise wave number $k_1$ in place
of $k$, and with the effective surface-scattering potential taken
in the form

\begin{equation}\label{SD-SM-V}
V(x)=\frac{2}{\pi}\,\left(\frac{\pi}{d}\right)^3 \,\xi(x).
\end{equation}
Thus, the surface scattering in a single-mode guiding structure
should manifest the effect of Anderson localization. In
particular, the average transmittance $\langle
T(L/L_{loc})\rangle$ is described by the universal expressions
(\ref{<1D-Ts>}) -- (\ref{1D-Tloc}) and, therefore, the transport
properties in such single-mode waveguides are governed by the
one-parameter scaling. For large localization length,
$L_{loc}/L\gg1$, the system reveals the ballistic transport
because its average transmittance $\langle
T(L/L_{loc})\rangle\approx1$, see equation~(\ref{1D-Tbal}). In the
other limit, $L_{loc}(k_n)/L\ll1$, the transmittance $\langle
T(L/L_{loc})\rangle$ is exponentially small in line with
equation~(\ref{1D-Tloc}). This implies strong electron/wave
localization.

Taking into account the specific form (\ref{SD-SM-V}) of the
surface-scattering potential and the correlation properties
(\ref{SD-Ksi}) for the corrugated surface profile $\xi(x)$, from
the general expression (\ref{1D-Lloc}) one can derive the
following explicit formula for the surface-scattering localization
length~\cite{MakYur89FMYu90},

\begin{equation}\label{SD-SM-Lloc}
L_{loc}^{-1}(k_1)=\frac{2\sigma^2}{\pi^2}
\left(\frac{\pi}{d}\right)^6\frac{W(2k_1)}{(2k_1)^2}.
\end{equation}
Since the potential (\ref{SD-SM-V}) is entirely determined by the
rough surface profile $\xi(x)$, the surface-scattering
localization length (\ref{SD-SM-Lloc}) is specified by the
roughness power spectrum $W(k_x)$. Therefore, by a proper
fabrication of a random surface profile $\xi(x)$ with specific
long-range correlations, one can arrange a desirable anomalous
ballistic transport within a given window inside the allowed
single-mode region (\ref{SM-k1}).

The rough surfaces with prescribed two-point correlations can be
constructed with the use of direct generalization of the method
discussed in Section 2 (for details, see also the
reference~\cite{Rice54}). Then the resulting surface profile
$\xi(x)$ can be obtained due to the following expression,

\begin{equation}\label{xi-beta}
\xi(x)=\sigma\,\int_{-\infty}^\infty\,dx'\,Z(x-x')\,\beta(x').
\end{equation}
Here the function $\beta(x)$ is determined by the power spectrum
$W(k_x)$ via the relation (\ref{beta-def}), and a delta-correlated
random process $Z(x)$ has standard statistical properties
(\ref{Zcorr}).

Below we demonstrate a possibility of constructing the
surface-disordered structures with anomalous ballistic transport
by considering two simple examples of a long-range correlated
surface profile.

{\bf(a)} Let us first consider the waveguide which is
non-transparent if the wave number $k_1$ is less than some value
$k_c$, and completely transparent for $k_1>k_c$. Evidently, such a
behavior can be observed if the transition point (mobility edge)
$k_1=k_c$ is located inside the allowed single-mode interval
(\ref{SM-k1}),
\begin{equation}\label{SM-kc}
0<k_cd/\pi<\sqrt{3}.
\end{equation}
In this case one can get the following expressions for the binary
correlator ${\cal W}(x)$ and power spectrum $W(k_x)$,
\numparts
\begin{eqnarray}
{\cal W}_1(x)&=&\frac{\sin(2k_cx)}{2k_cx},
\label{SM-W1}\\[6pt]
W_1(k_x)&=&\frac{\pi}{2k_c}\,\Theta(2k_c-|k_x|).\label{SM-FT1}
\end{eqnarray}
\endnumparts
With the use of the expressions (\ref{xi-beta}), (\ref{beta-def})
and (\ref{Zcorr}), the corrugated surface profile with the above
correlation properties can be shown to be described by the
function
\begin{equation}\label{SM-xi1}
\xi_1(x)=\frac{\sigma}{\sqrt{2\pi k_c}}
\int_{-\infty}^{\infty}dx'Z(x-x')\frac{\sin(2k_cx')}{x'}.
\end{equation}
Correspondingly, the inverse surface-scattering localization
length is of the {\it step-down} form,
\begin{equation}\label{SM-Lloc1}
L_{loc1}^{-1}(k_1)=\frac{\sigma^2}{4\pi k_c}
\left(\frac{\pi}{d}\right)^6 \frac{\Theta(k_c-k_1)}{k_1^2}.
\end{equation}
Therefore, as the wave number $k_1$ increases, the localization
length $L_{loc1}(k_1)$ also smoothly increases and then diverges
at $k_1=k_c$. Thus, within the region $k_1<k_c$ the average
transmittance $\langle T(L/L_{loc})\rangle$ is expected to be
exponentially small (\ref{1D-Tloc}). The condition for a strong
localization to the left from the mobility edge $k_1=k_c$ reads as
\begin{equation}\label{SM-loc1}
\frac{L}{L_{loc1}(k_c)}=\frac{\sigma^2L}{4\pi k_c^3}\,
\left(\frac{\pi}{d}\right)^6\gg1.
\end{equation}
In the interval $k_c<k_1<\pi\sqrt{3}/d$ a ballistic regime occurs
with a perfect transparency, $\langle T(L/L_{loc})\rangle=1$.

{\bf(b)} Second example refers to a complimentary situation when
for $k_1<k_c$ the waveguide is perfectly transparent, and for
$k_1>k_c$ is non-transparent. The corresponding expressions for
the correlator ${\cal W}(x)$ and its Fourier transform $W(k_x)$
are given by
\numparts
\begin{eqnarray}
{\cal W}_2(x)&=&\pi\delta(2k_cx)-\frac{\sin(2k_cx)}{2k_cx},
\label{SM-W2}\\[6pt]
W_2(k_x)&=&\frac{\pi}{2k_c}\,\Theta(|k_x|-2k_c).\label{SM-FT2}
\end{eqnarray}
\endnumparts
In this case the corrugated surface is described by a
superposition of a white noise and the roughness of the first
type,
\begin{equation}\label{SM-xi2}
\xi_2(x)=\frac{\sigma}{\sqrt{2\pi k_c}}\left[\pi Z(x)-
\int_{-\infty}^{\infty}dx'Z(x-x')\frac{\sin(2k_cx')}{x'}\right].
\end{equation}
Correspondingly, the inverse localization length is expressed by
the step-up function,
\begin{equation}\label{SM-Lloc2}
L_{loc2}^{-1}(k_1)=\frac{\sigma^2}{4\pi
k_c}\left(\frac{\pi}{d}\right)^6 \frac{\Theta(k_1-k_c)}{k_1^2}.
\end{equation}
As a consequence, in contrast with the first case, here the
surface-scattering localization length $L_{loc2}(k_1)$ diverges
below the mobility edge $k_1=k_c$. At this point $L_{loc2}(k_1)$
sharply falls down to the finite value $L_{loc2}(k_c+0)$ and then
smoothly increases with a further increase of $k_1$. In order to
observe the localization regime within the whole region
$k_c<k_1<\pi\sqrt{3}/d$ of finite values of $L_{loc2}(k_1)$, one
should assume that a strong localization is retained at upper
point $k_1=\pi\sqrt{3}/d$ of the single-mode region (\ref{SM-k1}),
\begin{equation}\label{SM-loc2}
\frac{L}{L_{loc2}(\pi\sqrt{3}/d)}=\frac{\sigma^2L}{12\pi k_c}\,
\left(\frac{\pi}{d}\right)^4\gg1.
\end{equation}
Therefore, in this second example the ballistic transport is
abruptly replaced by a strong localization at the mobility edge
point $k_1=k_c$.

Below, we demonstrate the above predictions by a direct numerical
simulation. For this, the inverse localization length
$L_{loc}^{-1}$ was computed with the use of the transfer matrix
method. Specifically, the continuous scattering potential
(\ref{SD-SM-V}) was approximated by the sum of delta kicks with
the spacing $\delta$ chosen much smaller than any physical length
scale in the model. Thus, the discrete analogue of the 1D wave
equation (\ref{1DSchr-eq}) was under consideration, with the
lengthwise wave number $k_1$ in place of $k$, and with the surface
scattering potential $V(x)$ in the expression (\ref{SD-SM-V}). In
this way the Shr\"odinger/wave equation was expressed in the form
of a two-dimensional Hamiltonian map describing the dynamics of a
classical particle under the external noise determined by the
scattering potential $V(x)$. As a result, the analysis of the
localization length was reduced to the computation of the Lyapunov
exponent $L_{loc}^{-1}$ associated with this map (see details in
references~\cite{IzKr99,IzKrUll01}).

\begin{figure}[htb]
\vspace{0.cm} \hspace{0.cm}
\begin{center}
\hspace{-0.5cm}
\includegraphics[width=2.6in,height=1.9in,angle=0]{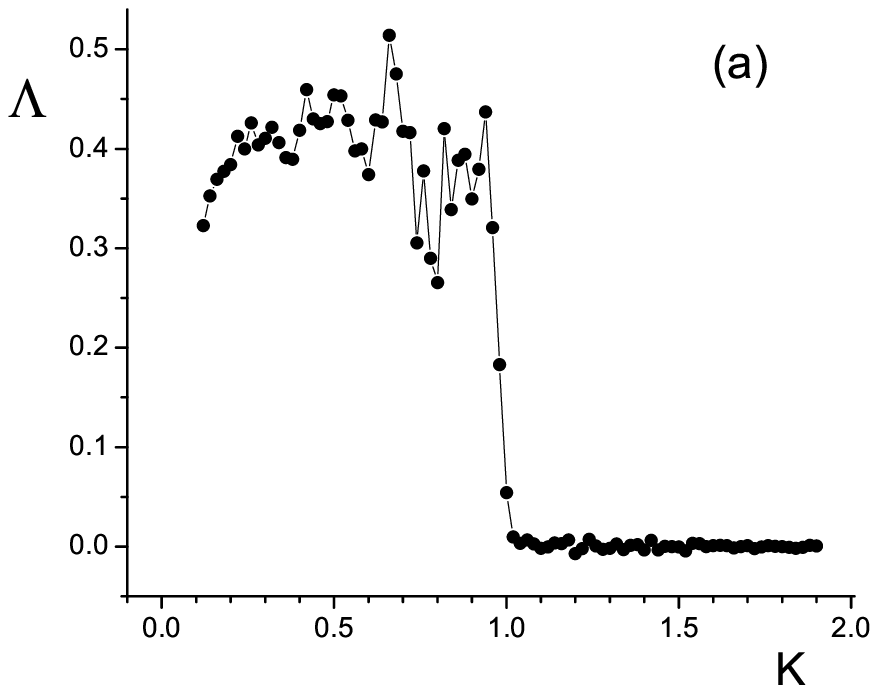}
\hspace{0cm}
\includegraphics[width=2.6in,height=1.9in,angle=0]{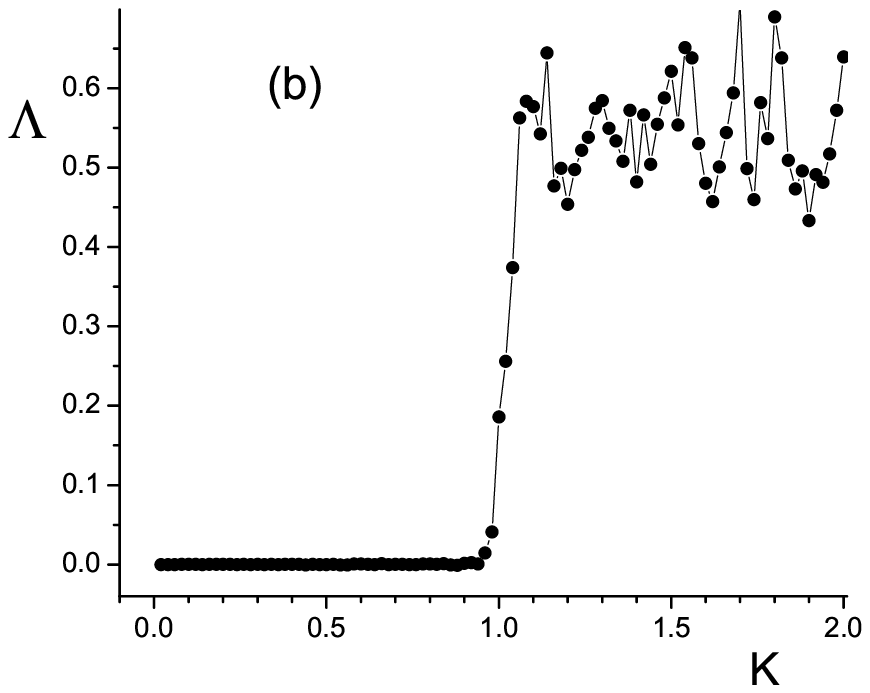}
\vspace{-0.5cm} \caption{\label{Fig3ab-SMW}Selective dependence of
the rescaled Lyapunov exponent on the wave number for two
realizations of a random surface with specific long-range
correlations. To show the main effect of the correlations, the
complimentary dependence of the Lyapunov exponent on $K$ is
demonstrated: (a) eigenstates are localized for $K<1$ and
delocalized for $K>1$; (b) wise-versa, complete delocalization for
$K<1$ alternates by the localization for $K>1$.} \vspace{0.0cm}
\end{center}
\end{figure}

Numerical data reported in figure~\ref{Fig3ab-SMW}, represent the
dependence of the dimensionless Lyapunov exponent
$\Lambda=c_0L_{loc}^{-1}$ on the normalized wave number
$K=k_1/k_c$ in the range $0<K<2$ corresponding to the single-mode
interval. The normalization coefficient $c_0$ was set to have
$\Lambda=K^{-2}$ for the delta-correlated potential. Two surface
profiles $\xi(x)$ were generated according to discrete versions of
the expressions~(\ref{SM-xi1}) and (\ref{SM-xi2}) determining
complimentary step-wise dependencies of the localization length
$L_{loc}(k_1)$ described by equations~(\ref{SM-Lloc1}) and
(\ref{SM-Lloc2}).

One can clearly see a nontrivial dependence of $\Lambda$ on the
wave vector $K$, which is due to specific long-range correlations
in $\xi(x)$. The data display sharp dependencies of $\Lambda$ on
$K$ when crossing the point $K=1$. Thus, by taking the size $L$ of
the scattering region in accordance with the
requirements~(\ref{SM-loc1}) or (\ref{SM-loc2}), one can arrange
anomalous transport in the single-mode guiding structure as is
predicted by the analytical theory.

\subsection{Multimode waveguide}

Now we examine the correlated surface scattering in {\it
multimode} guiding structure with the large total number $N_d>1$
of conducting channels, see expression (\ref{Nd}). According to
Landauer's concept \cite{Land92}, the {\it total average
transmittance} $\langle T\rangle$ of any quasi-1D conductor can be
expressed as a sum of {\it partial average transmittances}
$\langle T_n\rangle$ that describe the transport for every $n$-th
propagating normal mode,

\begin{equation}\label{SD-MM-T}
\langle T\rangle=\sum_{n=1}^{N_d}\langle T_n\rangle.
\end{equation}

From general theory of quasi-1D scattering systems it follows that
the transmission properties of any $n$-th conducting channel
($1\leq n\leq N_d$) are determined by two {\it attenuation
lengths}, the length $L_n^{(f)}$ of forward scattering and the
backscattering length $L_n^{(b)}$. For quasi-1D waveguides with
surface disorder the inverse scattering lengths are given
by~\cite{BFb79},
\begin{equation}\label{SD-MM-Lnf}
\frac{1}{L_n^{(f)}}=\sigma^2\frac{(\pi n/d)^2}{k_nd}
\sum_{n'=1}^{N_d}\frac{(\pi n'/d)^2}{k_{n'}d}W(k_n-k_{n'}),
\end{equation}
\begin{equation}\label{SD-MM-Lnb}
\frac{1}{L_n^{(b)}}=\sigma^2\frac{(\pi n/d)^2}{k_nd}
\sum_{n'=1}^{N_d}\frac{(\pi n'/d)^2}{k_{n'}d}W(k_n+k_{n'}).
\end{equation}
Here the lengthwise wave number $k_n$ is defined by the expression
(\ref{kn}). The results (\ref{SD-MM-Lnf}) and (\ref{SD-MM-Lnb})
can be obtained for the boundary-value problem (\ref{2D-BVP}) by
the diagrammatic Green's function method \cite{BFb79}, as well as
by the technique developed in the reference~\cite{McGM84}. Also,
these expressions can be derived by using the invariant imbedding
method extended to quasi-1D structures \cite{SFYM9899}. Note that
in a single-mode waveguide with $N_d=1$ the sum over $n'$ contains
only one term with $n'=n=1$. Therefore, in this case the
backscattering length $L_1^{(b)}$ is two times less then the
single-mode localization length (\ref{SD-SM-Lloc}),
$L_1^{(b)}=L_{loc}(k_1)/2$.

The expressions (\ref{SD-MM-Lnf}) and (\ref{SD-MM-Lnb}) manifest
that, in general, both attenuation lengths are contributed by
scattering of a given $n$-th propagating mode into all other
modes. This is the case when, for example, a rough surface profile
is either delta-correlated random process (``white noise'') with
constant power spectrum, $W(k_x)=const$, or has fast decreasing
binary correlator ${\cal W}(x)$ (or, the same, slowly decreasing
roughness power spectrum $W(k_x)$). Therefore, the quasi-1D
systems reveal three typical transport regimes, the regimes of a
{\it ballistic}, {\it diffusive} (metallic), and {\it localized
transport}.

Another peculiarity is that the expressions (\ref{SD-MM-Lnf}) and
(\ref{SD-MM-Lnb}) display rather strong dependence on the mode
index $n$, namely, the larger the number $n$ the smaller the
corresponding mode scattering lengths and as a consequence, the
stronger is the scattering of this mode into the others. As a
result, there is the following hierarchy of mode scattering
lengths,
\begin{equation}\label{SD-MM-Ln-hierarchy}
L_{N_d}^{(f,b)}<L_{N_d-1}^{(f,b)}<...<L_2^{(f,b)}<L_1^{(f,b)}.
\end{equation}
The smallest mode attenuation lengths, $L_{N_d}^{(f)}$ and
$L_{N_d}^{(b)}$, belong to the highest (last) channel with the
mode index $n=N_d$, while the largest scattering lengths,
$L_1^{(f)}$ and $L_1^{(b)}$, corresponds to the lowest (first)
channel with $n=1$. As was shown in the articles~\cite{SFYM9899},
due to this hierarchy of scattering lengths, even in the absence
of correlations in the corrugated surface $\xi(x)$, a very
important phenomenon of the {\it coexistence} of ballistic,
diffusive, and localized transport arises. Specifically, while
lowest modes can be in the ballistic regime, the intermediate and
highest modes can exhibit the diffusive and localized behavior,
respectively.  This effect seems to be generic for the transport
through the waveguides with random surfaces.

One can see now that unlike the single-mode case, the concept of
one-parameter scaling is no more valid for the transport in
multi-mode systems. There are two points that should be stressed
in this respect. On the one hand, the average partial
transmittances $\langle T_n\rangle$ entering the equation
(\ref{SD-MM-T}), are very different for different conducting
channels. On the other hand, and what is even more important, all
propagating modes turn out to be mixed due to the inter-mode
transitions. Therefore, the transmittance $\langle T_n\rangle$ of
a given $n$-th mode depends on the scattering into all modes.

From this analysis one can conclude that for multi-mode structures
with surface disorder the role of specific long-range correlations
should be much more sophisticated in comparison with that
discussed above for single-mode waveguides. First, such
correlations should result in the suppression of the interaction
between different propagating modes. This non-trivial fact turns
out to be crucial for the reduction of a system of mixed channels
with quasi-1D transport, to the coset of independent waveguide
modes with a purely 1D transport. Second, the same correlations
can provide a complete transparency of each independent channel,
similar to what happens in strictly 1D geometry.

To demonstrate these effects, let us take a random surface profile
$\xi(x)$ of the form
\begin{equation}\label{SD-MM-xi}
\xi(x)=\frac{\sigma}{\sqrt{\pi k_c}}\,\int_{-\infty}^\infty\,dx'\,
Z(x-x')\,\frac{\sin(k_cx')}{x'}\,,
\end{equation}
with the slowly decaying (in average) binary correlator which
results in the ``window'' function for the roughness power
spectrum,
\numparts
\begin{eqnarray}
{\cal W}(x)&=&\frac{\sin(k_cx)}{k_cx}\,,
\label{SD-MM-W}\\[6pt]
W(k_x)&=&\frac{\pi}{k_c}\,\Theta(k_c-|k_x|),\qquad\qquad k_c>0.
\label{SD-MM-RPS}
\end{eqnarray}
\endnumparts

From equations (\ref{SD-MM-Lnf}) and (\ref{SD-MM-Lnb}) one can see
that in the case of long-range correlations in a disordered
surface (\ref{SD-MM-xi}), the number of modes into which a given
$n$-th mode is scattered, i.e. the actual number of summands in
the equations (\ref{SD-MM-Lnf}) and (\ref{SD-MM-Lnb}), is entirely
determined by the width $k_c$ of the rectangular power spectrum
(\ref{SD-MM-RPS}). It is clear that if the distance
$|k_n-k_{n\pm1}|$ between neighboring quantum values of $k_n$ is
larger than the correlation width $k_c$,
\begin{equation}\label{SD-MM-cond}
|k_n-k_{n\pm1}|>k_c\,,
\end{equation}
then the transitions between all propagating modes are forbidden.
As a consequence, the sum over $n'$ in the
expression~(\ref{SD-MM-Lnf}) for the inverse forward scattering
length contains only diagonal term with $n'=n$ which describes
direct intra-mode scattering {\it inside} the channels. Moreover,
each term in the sum of equation~(\ref{SD-MM-Lnb}) for the inverse
backscattering length is equal to zero. As a result, the following
interesting properties arise.

{\bf(i)} All {\it high} propagating modes with indices $n$ that
satisfy the condition (\ref{SD-MM-cond}), turn out to be
independent of the others in spite of the interaction with rough
surface. Therefore, they form a coset of 1D non-interacting
conducting channels with finite length of forward scattering
$L_n^{(f)}$ and infinite backscattering length $L_n^{(b)}$,
\begin{equation}\label{SD-MM-Lnfb-CD}
\frac{1}{L_n^{(f)}}=\frac{\pi\sigma^2}{k_c}\,\frac{(\pi
n/d)^4}{(k_nd)^2}\,, \qquad\qquad\frac{1}{L_n^{(b)}}=0.
\end{equation}

{\bf(ii)} As is well known from the standard theory of 1D
localization (see, e.g., \cite{Ber73AR78,LGPb88,MT9801}), the
transport through any 1D disordered system is determined only by
the backscattering length $L_n^{(b)}$ and does not depend on the
forward-scattering length $L_n^{(f)}$. Since the former diverges
for every independent channel in line with the expression
(\ref{SD-MM-Lnfb-CD}), all of them exhibit the ballistic transport
with the partial average transmittance $\langle T_n\rangle=1$.
This means that according to the Landauer's formula
(\ref{SD-MM-T}), the transmittance of the coset of such
independent ballistic modes is simply equal to their total number.

{\bf(iii)} As for {\it low} propagating modes with the indices $n$
contradicting to the condition (\ref{SD-MM-cond}), they remain to
be mixed by surface scattering because the roughness power
spectrum (\ref{SD-MM-RPS}) is non-zero for them,
$W(k_n-k_{n'})=\pi/k_c$. These {\it mixed modes} have finite
forward- and back-scattering lengths. Therefore, for large enough
waveguide length $L$, they do not contribute to the total
transmittance $\langle T\rangle$ and the latter is equal to the
number of independent ballistic modes.

Note that the distance $|k_n-k_{n\pm1}|$ between neighboring
quantum wave numbers $k_n$ and $k_{n\pm1}$ increases as the mode
index $n$ increases. Therefore, the inequality (\ref{SD-MM-cond})
restricts the mode index $n$ from below. That is why, in contrast
with the conventional situation associated with the hierarchy of
mode scattering lengths (\ref{SD-MM-Ln-hierarchy}), the low modes
are mixed and non-transparent, while high propagating modes are
independent and ballistic. Because of the sharp behavior of the
roughness power spectrum (\ref{SD-MM-RPS}), the transition from
mixed to independent modes is also sharp.

More analytical results can be obtained for waveguides with large
number of conducting channels (\ref{Nd-large}), if the quantum
numbers $n$ of independent ballistic modes are also large,
$N_d\approx kd/\pi\geq n\gg1$. In this case the inequality
(\ref{SD-MM-cond}) is reduced to the requirement $|\partial
k_n/\partial n|>k_c$ which can be rewritten in the following
explicit form
\begin{equation}\label{SD-MM-Nmix}
n>N_{mix}=\left[\frac{(kd/\pi)}{\sqrt{1+(k_cd/\pi)^{-2}}}\right].
\end{equation}
We remind that square brackets stand for the integer part of the
inner expression.

The condition (\ref{SD-MM-Nmix}) determines the total number
$N_{mix}$ of mixed non-transparent modes, the total number
$N_{bal}=N_d-N_{mix}$ of independent ballistic modes, and the
critical value of the mode index $n$ that divides these two
groups. All propagating modes with $n>N_{mix}$ are independent and
fully transparent, otherwise, they are mixed for $n\leq N_{mix}$
and characterized by finite scattering lengths $L_n^{(f)}$ and
$L_n^{(b)}$. Therefore, the total average transmittance
(\ref{SD-MM-T}) of the multi-mode structure is
\begin{equation}\label{SD-MM-Tcd}
\langle T\rangle=[kd/\pi]-\left[(kd/\pi)/\alpha_c\right],\qquad
\alpha_c=\sqrt{1+(k_cd/\pi)^{-2}}\,.
\end{equation}

The numbers $N_{mix}$ and $N_{bal}$ of mixed non-transparent and
independent ballistic modes are governed by two parameters, the
mode parameter $kd/\pi$ and the dimensionless correlation
parameter $k_cd/\pi$. In the case of ``weak" correlations when
$k_cd/\pi\gg1$, the number of mixed modes $N_{mix}$ is of the
order of $N_d$,
\begin{equation}\label{SD-MM-Nmix-wc}
N_{mix}\approx\left[\left(\frac{kd}{\pi}\right)-
\frac{1}{2}\left(\frac{kd}{\pi}\right)
\left(\frac{k_cd}{\pi}\right)^{-2}\right] \quad\mbox{for}\quad
k_cd/\pi\gg1.
\end{equation}
Consequently, in this case the number of ballistic modes $N_{bal}$
is small, or there are no such modes at all. If the parameter
$k_cd/\pi$ tends to infinity, $k_cd/\pi\to\infty$, the rough
surface profile becomes white-noise-like and, naturally,
$N_{mix}\to N_d$.

The most appropriate case is when a random surface profile is
strongly correlated so that the correlation parameter is small,
$k_cd/\pi\ll1$. Then the number of mixed non-transparent modes
$N_{mix}$ is much less than the total number of propagating modes
$N_d$,
\begin{equation}\label{SD-MM-Nmix-sc}
N_{mix}\approx\left[\left(\frac{kd}{\pi}\right)
\left(\frac{k_cd}{\pi}\right)\right]\ll N_d \quad\mbox{for}\quad
k_cd/\pi\ll1.
\end{equation}
Therefore, the number of independent modes $N_{bal}$ is large.
When the correlation parameter $k_cd/\pi$ decreases and becomes
anomalously small, $k_cd/\pi<(kd/\pi)^{-1}\ll1$, the number of
mixed modes $N_{mix}$ vanishes and {\it all modes} become
independent and perfectly transparent. Evidently, if the
correlation parameter $k_cd/\pi$ vanishes, $k_cd/\pi\to0$, the
roughness power spectrum (\ref{SD-MM-RPS}), becomes
delta-function-like and, as a consequence, $N_{mix}=0$. In this
case the correlated disorder results in a perfect transmission of
electrons or waves.

Finally, let us briefly discuss the expression (\ref{SD-MM-Tcd})
for the total average transmittance. In figure 3 an unusual
step-wise dependence of $\langle T\rangle$ on the mode parameter
$kd/\pi$ is shown, that is governed by the width $k_c$ of the
rectangular power spectrum (\ref{SD-MM-RPS}).
\begin{figure}[htb]
\vspace{0.cm} \hspace{4.0cm}
\begin{center}
\includegraphics[width=4.2in,height=2.4in,angle=0]{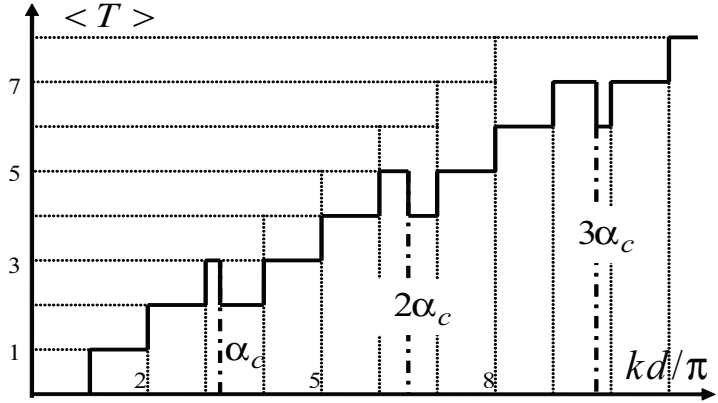}
\vspace{0.0cm} \caption{Step-wise transmittance (\ref{SD-MM-Tcd})
of surface-disordered guiding system versus the mode parameter
$kd/\pi$. In the figure, the value of the normalized correlation
parameter $k_cd/\pi=0.32$. }
\end{center}
\end{figure}

Let us discuss this figure. Within the region $kd/\pi<\alpha_c$,
the inter-mode transitions caused by the specific surface
correlations, are forbidden for all conducting channels.
Therefore, all propagating modes are independent and ballistic,
and the second term (the number of mixed modes) in the expression
(\ref{SD-MM-Tcd}) for the total transmittance is equal to zero.
Here the transmittance exhibits a ballistic step-wise increase
with an increase of the parameter $kd/\pi$. Each step up arises
for an integer value of the mode parameter $kd/\pi$, when a new
conducting channel emerges in the guiding structure.

Otherwise, when $kd/\pi\geq\alpha_c$, in addition to the standard
steps up originated from the first term in equation
(\ref{SD-MM-Tcd}), there are also the steps down connected with
the second term. These steps down are provided by the correlated
surface scattering and arise when a successive low mode becomes
mixed and non-transparent. In other words, the positions of the
$n$-th steps down are at the mobility edge point
$kd/\pi=n\alpha_c$ where the $n$-th conducting channels closes.
Since the values of the ratio $kd/\pi\alpha_c$ are determined by
the correlation parameter $k_c$, the positions of steps down, in
general, do not coincide with those of steps up. The situation may
also occur when the steps up and down cancel each other within
some interval of the mode parameter $kd/\pi$. The experimental
observation of the discuseed non-coventional dependence seems to
be highly interesting.


\ack This research was partially supported by the CONACYT
(M\'exico) grant No~43730, and by the VIEP-BUAP (M\'exico) under
the grant 5/G/ING/05.

\end{document}